\documentclass[aps,showpacs,preprintnumbers,amsmath,amssymb,nofootinbibt,twocolumn]{revtex4}
\usepackage{graphicx}
\usepackage{amsmath}
\usepackage{amssymb}
\usepackage{amsfonts}
\usepackage{bm}
\usepackage{color}

\def\be{\begin{equation}}
\def\ee{\end{equation}}
\def\bea{\begin{eqnarray}}
\def\eea{\end{eqnarray}}

\begin{document}
%\preprint{gr-qc/yymmnnn}

\title{Irreversible thermodynamic description of interacting dark energy - dark matter cosmological models}

\author{Tiberiu Harko$^1$}\email{t.harko@ucl.ac.uk}
\author{Francisco S.N.~Lobo$^{2}$}\email{flobo@cii.fc.ul.pt}
\affiliation{$^1$Department of Mathematics, University College London, Gower Street, London WC1E 6BT, United Kingdom}
\affiliation{$^2$Centro de Astronomia e Astrof\'{\i}sica da Universidade de Lisboa, Campo Grande, Ed. C8 1749-016 Lisboa, Portugal}

\date{\today}

\begin{abstract}

We investigate the interaction between dark energy and dark matter in the framework of irreversible thermodynamics of open systems with matter creation/annihilation.  We consider dark energy and dark matter as an interacting  two component (scalar field and ``ordinary'' dark matter) cosmological fluid in a homogeneous spatially flat and isotropic Friedmann-Robertson-Walker (FRW) Universe. The thermodynamics of open systems as applied together with the gravitational field equations to the two component cosmological fluid leads to a generalisation of the elementary dark
energy-dark mater interaction theory, in which the decay (creation) pressures are explicitly considered as parts of the cosmological fluid stress-energy tensor. Specific models describing coherently oscillating scalar waves, leading to a high particle production at the beginning of the oscillatory
period, and models with a constant potential energy scalar field are considered. Furthermore, exact and numerical solutions of the gravitational field equations with dark energy-dark matter interaction are also obtained.

\end{abstract}


\maketitle

\section{Introduction}

The observational data from type Ia supernovae, initially reported in \cite{Ri98}, have generated a large theoretical and observational effort for the understanding  of the observed present accelerated expansion of the Universe. Subsequent work on type Ia supernovae \cite{Hi}, the cosmic microwave
background \cite{Ko}, and baryon acoustic oscillations \cite{Perc} fully support the initial interpretation of the observational data that the expansion of the
Universe is accelerating. The late-time cosmic acceleration is usually assumed to be driven by a fluid/field generically denoted dark energy \cite{PeRa03}. Presently very little is known about dark energy, namely, its possible composition or its  structure. Two main scenarios
have been proposed to explain the nature of the dark energy: a cosmological constant $\Lambda $ \cite{PeRa03},  or a scalar field, usually called quintessence \cite{quint}. The action for gravity and the scalar field is $S=\int{\left[R/16\pi G -(1/2)\nabla ^{\alpha }\phi \nabla _{\alpha }\phi -V(\phi )\right]\sqrt{-g}d^4x}$, where $V(\phi )$ is the self-interaction potential \cite{Fa04}.

Another of the central issues in modern astrophysics is the dark matter problem (see \cite{Sal} for an extensive review of the recent results
of the search for dark matter). The necessity of considering the existence of dark matter at a galactic and extragalactic scale is required by two fundamental observational evidences: the behavior of the galactic rotation curves,
and the mass discrepancy in clusters of galaxies, respectively. On the galactic/intergalactic scale the rotation curves
of spiral galaxies \cite{Bin, Per, Bor} provide compelling evidences pointing towards the problems Newtonian gravity and/or standard general relativity has to face at these scales. The behavior of the galactic rotation curves and of the virial mass of galaxy clusters is usually explained by
postulating the existence of some dark (invisible) matter, distributed in a spherical halo around the galaxies. The
dark matter is assumed to be a cold, pressure-less medium. Many possible candidates for dark matter have been
proposed, the most popular ones being the WIMPs (Weakly Interacting Massive Particles) (for a review of the
particle physics aspects of the dark matter see \cite{Ov}). While extremely small, their interaction cross sections with
normal baryonic matter, are expected to be non-zero, so that their direct experimental detection may be possible.

In this context, cosmological evolution and dynamics are largely dominated by dark energy and dark matter. Dark energy has a repulsive effect, driving the Universe to accelerate, while dark matter is gravitationally attractive. In the standard approach to cosmology there is no interaction between these two components. Since the gravitational effects of the dark energy and of the dark matter are opposite (i.e., gravitational repulsion versus gravitational attraction) and since dark energy is very homogeneously distributed, while dark matter clumps around ordinary matter, one expects that any dynamic interaction between these two dark components of the Universe would be extremely weak, or even negligible. However, the possibility of such an interaction cannot be excluded {\it a priori} and, following some early proposals \cite{int}, presently interacting dark matter-dark energy models were extensively investigated in the literature \cite{lit}. In the standard approach, one may model dark energy as a scalar field with energy density $\rho_{\phi }$ and pressure $p_{\phi }$, while dark matter is described as a matter fluid with density $\rho _{DM}$ and pressure $p_{DM}$, satisfying an equation of state $w_{DM}=p_{DM}/\rho_{DM}\equiv 0$. By assuming a spatially flat Friedman-Robertson-Walker (FRW)
background with scale factor, $a(t)$, and allowing for creation/annihilation between the dark energy (scalar field) and the
dark matter fluid at a rate $Q$, the equations describing the variations of the dark energy and dark matter densities $\rho _{\phi }$  and $\rho _{DM}$ can be written as \cite{int,lit}
\bea\label{1}
\dot{\rho _{\phi }}+3H\left(1+w_{\phi}\right)\rho _{\phi}&=&-Q,
    \\
%\ee
%\be
\label{2}
\dot{\rho _{DM}}+3H\left(1+w_{DM}\right)\rho _{DM}&=&+Q,
\eea
respectively, where $H \equiv \dot{a}/a$ is the Hubble function. The derivatives with respect to the cosmological time, $t$, will be indicated in the following by an overdot.
The dark energy  and the dark matter create/decay into one another via the common creation/annihilation rate $\pm Q$. Hence $Q$ describes the interaction between the two dark components of the Universe.  When $Q > 0$, dark energy is converted to dark matter, while if $Q < 0$, dark matter is converted to dark
energy. Since there is no fundamental theoretical approach that may specify the functional form of the coupling between dark energy and dark matter,
presently coupling models are  necessarily phenomenological, although one might view some couplings more physical or more natural than others. Hence a large number of functional forms for $Q$ have been proposed, and investigated in the literature, such as $Q=\rho _{crit}^0(1+z)^3H(z)I_Q(z)$, where $z$ is the redshift, and $I_Q(z)$ is an interaction function that depends on the redshift \cite{Cueva}, or $Q\propto \rho _{DM}\dot{\phi }$, and $Q\propto H\rho _{DM}$, respectively \cite{Singl}.

In writing down Eqs.~(\ref{1}) and (\ref{2}), one assumes the idealised picture of a very quick decay in which the
final products reach equilibrium states immediately. However,  the dark energy/dark matter creation/annihilation period may be characterised
by complicated nonequilibrium processes,  with a highly nonequilibrium distribution of the produced particles, subsequently relaxing to an equilibrium state. Thermodynamical systems in which matter creation occurs fits in the class of open thermodynamical systems in which the usual adiabatic conservation
laws are modified thereby including irreversible matter creation \cite{Prig}. The thermodynamics of open systems were first applied to cosmology in \cite{Prig}. Explicit inclusion of the matter creation in the matter
stress-energy tensor in the Einstein field equations leads to a three stage cosmological history, starting from an instability of the vacuum. During the first stage the Universe is driven from an initial fluctuation of the vacuum to a de Sitter phase, and particle creation occur. The de Sitter phase does exist during the decay time of its constituents (second stage)
and ends, after a phase transition into the usual FRW Universe. The phenomenological approach of \cite{Prig} was further discussed and generalised in \cite{Cal} through a covariant formulation allowing specific entropy variation as
usually expected for non-equilibrium processes. Cosmological models involving
irreversible matter creation have been considered in \cite{cosm}.

It is the purpose of the present paper to apply the thermodynamics of open
systems as developed in \cite{Prig} and \cite{Cal} to a cosmological fluid mixture consisting of
two components: dark energy, described by a scalar field, and dark matter, modeled as an ordinary matter fluid,  in which particle decay and production
occur. This situation may be specific to both early and late stages of cosmological evolution.  The thermodynamics of irreversible processes as applied to
 cosmological models with interacting dark energy and dark matter leads to a self-consistent
description of the dark energy and dark matter particle creation/annihilation processes, which in turn determine the whole dynamics and future evolution of
the Universe.

The present paper is organised as follows. In Section~\ref{sect2} we present, in some detail, due to its important role, the thermodynamical theory of irreversible matter creation processes. The theory is applied to a
two-component cosmological fluid with interacting dark energy and dark matter,  and the resulting gravitational field equations are written
down in Section~\ref{sect3}. Particular models, and exact and numerical solutions to the field equations are considered in Section~\ref{sect4}. In Section~\ref{sect5} we discuss and conclude our results. Throughout the paper we use a system of units so that $8\pi G=c=1$.

\section{Thermodynamics of Irreversible Cosmological Matter Creation}\label{sect2}

We consider a cosmological volume element $V$ containing $N$ particles. For a closed system,  $N$ is constant, and the corresponding thermodynamic conservation of the internal energy $E$ is expressed by the first law of thermodynamics as \cite{Prig}
\be\label{encons}
dE=dQ-pdV,
\ee
where $dQ$ is the heat received by the system during time $dt$, $p$ is the thermodynamic pressure, and $V$ is any comoving volume. By introducing the energy density $\rho$ defined as $\rho = E/V$, the particle number density $n$ given by $n = N/V$, and the heat per unit particle $dq$, with $dq = dQ/N$, Eq.~(\ref{encons}) becomes
\be\label{encons1}
d\left(\frac{\rho }{n}\right)=dq-pd\left(\frac{1}{n}\right).
\ee

Equation (\ref{encons1}) is also valid for open systems in which $N$ is time dependent, $N=N(t)$.

\subsection{General relativistic covariant formulation of matter creation}

In a general-relativistic framework the basic macroscopic variables which describe the thermodynamic states of a relativistic simple fluid are the energy-momentum tensor $T_{\mu \nu}$, the particle flux vector $N^{\mu }$, and the entropy flux vector $s^{\mu }$. The energy-momentum tensor satisfies the conservation law $\nabla _{\nu }T^{\mu \nu}=0$. By taking into account matter creation the energy-momentum tensor can be written as
 \be
 T^{\mu \nu}=\left(\rho +p+p_c\right)u^{\mu }u^{\nu }-\left(p+p_c\right)g^{\mu \nu},
 \ee
 where the creation  pressure takes into account dissipative effects.

The particle flux vector is given by $N^{\mu} =nu^{\mu }$, where $n$ is the particle number density, and $u^{\mu }$ is the four-velocity of the fluid. The particle flux vector satisfied the balance equation
\be\label{n}
\nabla _{\mu}N^{\mu }=\Psi ,
\ee
where the function $\Psi $ is a particle source for $\Psi >0$, and a particle sink for $\Psi <0$. In standard cosmology $\Psi $ is usually assumed to be zero. We also introduce the entropy flux $s^{\mu }$, defined as $s^{\mu }=n\sigma u^{\mu }$ \cite{Cal}, where $\sigma $ is the specific entropy per particle. The second law of thermodynamics requires that $\nabla _{\mu }s^{\mu }\geq 0$.  For an open thermodynamic system with temperature $T$ in the presence of matter creation the Gibbs equation is
\be
nTd\sigma =d\rho -\frac{\rho +p}{n}dn.
\ee

By using the above equations one can immediately obtain the entropy balance equation as \cite{Cal}
\be\label{eq1}
\nabla _{\mu }s^{\mu }=-\frac{p_c \Theta }{T}-\frac{\mu \Psi}{T},
\ee
where $\Theta =\nabla _{\mu }u^{\mu}$ is the expansion of the fluid, and the chemical potential $\mu $ is given by Euler's relation $\mu =\left(\rho +p\right)/n-T\sigma$.

 In the following we consider  that the particles are created in the space-time in such a way that they are in thermal equilibrium with the already
existing ones. Then the entropy production is due only to the matter creation. Moreover, we shall assume for the creation pressure  $p_c $ the following phenomenological ansatz \cite{Prig, Cal}
\be
p_c =-\alpha \frac{\Psi}{\Theta},
\ee
where $\alpha >0$. With this choice we obtain for the entropy balance the equations
\be
\nabla _{\mu }s^{\mu }=\frac{\Psi}{T}\left(\alpha -\mu \right)=\Psi \sigma +\left(\alpha -\frac{\rho +p}{n}\right)\frac{\Psi}{T}=\Psi \sigma +n\dot{\sigma },
\ee
where $\dot{\sigma}=u^{\mu }\nabla _{\mu }\sigma =d\sigma /ds$, which, together with Eq.~(\ref{eq1}) gives for the specific entropy production the relation \cite{Cal}
\be\label{eq2}
\dot{\sigma }=\frac{\Psi}{nT}\left(\alpha -\frac{\rho +p}{n}\right).
\ee

If we constrain our formalism by requiring that the specific entropy per particle is constant, $\sigma ={\rm constant}$, then Eq.~(\ref{eq2}) fixes the form of $\alpha $ as $\alpha =\left(\rho +p\right)/n$, giving for the creation pressure the expression \cite{Cal}
\be\label{pc}
p_c=-\frac{\rho +p}{n\Theta }\Psi .
\ee

By taking into account the condition of the constancy of the specific entropy, the Gibbs equation becomes
\be\label{ad}
\dot{\rho }=\left(\rho +p\right)\frac{\dot{n}}{n}.
\ee

\subsection{Matter creation in homogeneous and isotropic cosmological models}

In the case of a homogeneous and isotropic space-time we adopt a comoving frame so that the components of the four-velocity are given by $u^{\mu }=\left(1,0,0,0\right)$. Moreover, we assume that the thermodynamic as well as the geometric parameters are a function of the time $t$ only. Then the derivative of any function $f(t)$ with respect to the line element $s$ coincides with the ordinary time derivative, $\dot{f}=u^{\mu }\nabla _{\mu }f=df/dt$. Moreover, the expansion of the fluid is given by $\nabla _{\mu }u^{\mu }=\dot{V}/V$.
Equation (\ref{ad}) can be written in a number of equivalent forms as
\be\label{rhodot}
\dot{\rho }=\left(\frac{h}{n}\right)\dot{n},
\ee
where  $h = \rho + p$ is the enthalpy (per unit volume) of the fluid, or, equivalently,
\be
p=\dot{\rho }-\rho \frac{\dot{n}}{n}.
\ee

The Einstein field equations
\be
R_{\mu \nu }-\frac{1}{2}g_{\mu \nu }R=T_{\mu \nu },
\ee
involve the macroscopic stress-energy tensor $T_{\mu \nu }$, which,  in
the cosmological case, corresponds to a perfect fluid. It is characterised by a phenomenological energy density $\rho $ and pressure $\bar{p}$, and its components are given by
\be
T_0^0=\rho, T_1^1=T_2^2=T_3^3=-\bar{p}.
\ee
In addition to the Einstein field equations we have the Bianchi identities, which lead to $\nabla _{\nu }T_{\mu }^{\nu }=0$,  and to the relation
\be\label{cons2}
d(\rho V ) = -\bar{p}dV.
\ee

In the presence of adiabatic irreversible matter creation the appropriate analysis
must be performed in the context of open systems. This involves the inclusion of a
supplementary creation/annihilation pressure $p_c$, as we may write Eq.~(\ref{ad}) in a form similar to Eq.~(\ref{cons2}), namely \cite{Prig}
\be\label{cons3}
d(\rho V ) = -\left(p+p_c\right)dV,
\ee
where from Eq.~(\ref{pc}) it follows that the creation pressure is given by
\be\label{pc}
p_c = -\left(\frac{h}{n}\right)\frac{d(nV)}{dV}=-\left(\frac{h}{n}\right)\frac{V}{\dot{V}}\left(\dot{n}+\frac{\dot{V}}{V}n\right).
\ee

Creation of matter corresponds to a (negative) supplementary pressure $p_c$, which must be considered as part of the cosmological pressure $\bar{p}$ entering into the Einstein field equations (decaying of matter leads to a positive decay pressure),
\be
\bar{p} = p + p_c.
\ee
The entropy change $dS$ in an open thermodynamic system can be decomposed into an entropy flow $d_0S$, and the entropy creation $d_iS$,
\be\label{ent1}
dS = d_0S + d_iS,
\ee
with $d_iS \geq 0$. To evaluate $dS$ we start from the total differential of the entropy,
\be\label{ent2}
Td(sV ) = d(\rho V ) + pdV -\mu d(nV ),
\ee
where $ s= S/V\geq 0$ and $\mu n = h - Ts$, $\mu \geq 0$ being the chemical potential. In a homogeneous system $d_0S = 0$, but matter creation contributes to the entropy production. From Eqs.~(\ref{ent1}) and (\ref{ent2}) we obtain \cite{Prig}
\be
T\frac{dS}{dt}= T\frac{d_iS}{dt}= T\frac{s}{n}\frac{d(nV)}{dt}.
\ee
To complete the problem we need one more relation between the particle number $n$ and $V$, describing the time – dynamics of $n$ as a result of matter creation (decay) processes. This relation is given by Eq.~(\ref{n}), which in the case of a homogeneous and isotropic cosmological model takes the form
\be\label{Psi}
\frac{1}{V}\frac{d(nV )}{dt}=\Psi (t),
\ee
where $\Psi(t)$ is the matter creation (or decay) rate ($\Psi (t) > 0$ corresponds to particle creation, while $\Psi (t) < 0$ corresponds to particle decay) \cite{Prig, Cal}. The creation pressure (\ref{pc}) depends on the matter creation (decay) rate, thereby coupling Eqs.~(\ref{pc}) and (\ref{Psi}) to each other and, although indirectly, both of them with the energy conservation law (\ref{cons3}), which is contained in the Einstein field equations themselves. The entropy production can also be expressed as a function of the matter creation rate as
\be
S(t)=S\left(t_0\right)e^{\int_{t_0}^{t}{\frac{\Psi (t)}{n}dt}}.
\ee

\section{Cosmological dynamics in a Universe with irreversible dark energy-dark matter interaction }\label{sect3}

We shall model the  Universe  as an open thermodynamical system, consisting of a two-component (dark energy and dark matter) perfect fluid, with the particle number densities denoted by $n_{\phi }$, and  $n_{DM}$, respectively. $n_{\phi }$ corresponds to the ``particles'' of the scalar field, while $n_{DM}$ is the particle number of the dark matter. We denote the corresponding energy densities by $\rho _{\phi }$ and $\rho _{DM}$, respectively. The stress-energy tensor of the two-component cosmological fluid is given by
\be
T_{\mu }^{\nu }=T_{\mu }^{(\phi )\nu }+T_{\mu }^{(DM)\nu }=\rho u_{\mu }u^{\nu }-\bar{p}\;\delta _{\mu }^{\nu },
\ee
where $u^{\mu }=dx^{\mu }/ds$ is the four-velocity, and
\be
\rho =\rho _{\phi }+\rho _{DM}, \qquad \bar{p}= \bar{p}_{\phi }+\bar{p}_{DM}.
\ee

The energy density and pressure of the dark energy are given by $\rho _{\phi }=\dot{\phi }^2/2+U(\phi )$ and $p_{\phi }=\dot{\phi }^2/2-U(\phi )$, respectively, where $U(\phi )$ is the self-interaction potential.  We suppose that neither the particle numbers nor the stress-energy of the components are separately conserved, that is, particle inter-conversion and exchange of energy and momentum between the two components are admitted. The cosmological
fluid mixture is characterised by a total energy density $\rho = \rho _{\phi } + \rho _{DM}$, total thermodynamic pressure $\bar{p} = \bar{p}_{\phi }
 + \bar{p}_{DM}$ and a total particle number $n = n_{\phi } + n_{DM}$.
We consider that the geometry of the spacetime is described by the flat FRW line element, given by
\be
ds^2=dt^2-a^2(t)\left(dx^2+dy^2+dz^2\right).
\ee
We shall assume that the particle number densities $n_{\phi }$ and $n_{DM}$ of each component
of the fluid obey the following balance laws,
\bea \label{rate1}
\dot{n}_{\phi}  + 3Hn_{\phi } &=& -\Gamma _1\rho _{\phi },
%\ee
%\be
   \\
\label{rate2}
\dot{n}_{DM}  + 3Hn_{DM } &=& \Gamma _2\rho _{\phi },
\eea
respectively, where $\Gamma _1\neq 0$ and $\Gamma _2\neq 0$ are arbitrary functions. Equations (\ref{rate1}) and (\ref{rate2}) describe the decay of the dark energy $\phi $-particles, and the creation of the dark matter particles, with a scalar field decay rate and a dark matter creation
rate $\Psi(t) \propto \rho _{\phi }$. Thus, the dynamics of $\phi $-particle decay and the creation of dark matter is governed in the present model by the scalar field via its energy density. From Eqs.~(\ref{rate1}) and (\ref{rate2}) it follows that the total particle number $n$ obeys the
balance equation
\be
\dot{n} + 3Hn = \left(\Gamma _2-\Gamma _1\right)\rho _{\phi }.
\ee
Hence in the case of an interacting dark matter and dark energy the total particle number conservation occurs only in very special cases, and therefore we shall suppose that generally $\Gamma _1\neq \Gamma _2$.

In the framework of the thermodynamics of irreversible processes particle creation
and decay gives rise to a decay and a creation thermodynamic pressure, given
by
\be
p_c^{(\phi )}=\frac{\Gamma _1\left(\rho _{\phi }+p_{\phi }\right)\rho _{\phi }}{3Hn_{\phi }},
\ee
and
\be
p_c^{(DM)}=-\frac{\Gamma _2\left(\rho _{DM}+p_{DM}\right)\rho _{\phi }}{3Hn_{DM}},
\ee
respectively, while the total creation pressure becomes
\bea
p_c^{(total)}&=&p_c^{(\phi )}+p_c^{(DM)}
   \nonumber\\
&=&\frac{\rho _{\phi }}{3H}\left[\frac{\Gamma _1\left(\rho _{\phi }+p_{\phi }\right)}{n_{\phi }}-\frac{\Gamma _2\left(\rho _{DM}+p_{DM}\right)}{n_{DM}}\right]. \nonumber\\
\eea

Using the results obtained above the complete Einstein gravitational field equations describing the dynamics of a flat FRW spacetime
filled with a mixture of interacting dark matter (scalar field) and dark matter can be expressed in the form
\bea
3H^2&=&\rho _{\phi }+\rho _{DM},
%\ee
%\bea
   \\
2\dot{H}+3H^2&=&-p_{\phi}-p_{DM}
    \nonumber\\
&&\hspace{-1.25cm}-\frac{\rho _{\phi }}{3H}\left[\frac{\Gamma _1\left(\rho _{\phi }+p_{\phi }\right)}{n_{\phi }}-\frac{\Gamma _2\left(\rho _{DM}+p_{DM}\right)}{n_{DM}}\right],\nonumber\\
\eea
and
where $\rho _{DM}=\rho _{DM}\left(n_{DM}\right)$ and $p_{DM}=p_{DM}\left(n_{DM}\right)$. The dynamical evolution of the dark energy and dark matter particles $n_{\phi}$ and $n_{DM}$ is given by Eqs.~(\ref{rate1}) and (\ref{rate2}), respectively, while the energy density and pressure of the dark energy is given by
\be
\rho _{\phi}=\frac{\dot{\phi}^2}{2}+U(\phi ),\qquad p_{\phi }=\frac{\dot{\phi}^2}{2}-U(\phi ),
\ee
where $U(\phi)$ is the scalar field self-interaction potential.

As applied to each component of the cosmological fluid,  Eq.~(\ref{rhodot}), the second law of thermodynamics for open systems provides the relationships
\be\label{scalf}
\dot{\rho }_{\phi }+3H\left(\rho _{\phi }+p_{\phi }\right)+\frac{\Gamma _1\left(\rho _{\phi }+p_{\phi }\right)\rho _{\phi }}{n_{\phi }}=0,
\ee
and
\be\label{dm}
\dot{\rho }_{DM }+3H\left(\rho _{DM }+p_{DM }\right)=\frac{\Gamma _2\left(\rho _{DM }+p_{DM }\right)\rho _{\phi }}{n_{DM }}\,,
\ee
respectively.

Equation (\ref{scalf}), which describes the dynamics of the dark energy during its interaction with dark matter, can be written in an equivalent form as
\be\label{scaleq}
\ddot{\phi }+3H\dot{\phi }+\Gamma \left(\phi, \dot{\phi },U\right)\dot{\phi }+U'(\phi )=0,
\ee
where we have denoted $\Gamma \left(\phi, \dot{\phi }, U\right)=\Gamma _1\rho _{\phi }/n_{\phi }$. Therefore in the framework of the thermodynamics of irreversible
processes a friction term in the scalar field Eq.~(\ref{scaleq}) arises naturally, and in a general form, as a direct consequence of the second law of thermodynamics as applied to an open system.

Adding Eqs.~(\ref{scalf}) and (\ref{dm}), the evolution of the total energy density $\rho  =
\rho _{\phi } + \rho _{DM}$ of the cosmological fluid is governed by the equation
\bea
&&\dot{\rho}+3H\left(\rho +p_{\phi }+p_{DM}\right)=\nonumber\\
&&\rho _{\phi }\left[\frac{\Gamma _2\left(\rho _{DM}+p_{DM}\right)}{n_{DM}}-\frac{\Gamma _1\left(\rho _{\phi }+p_{\phi }\right)}{n_{\phi }}\right].
\eea

For the entropy of the newly created matter we obtain
\be\label{42}
S_{DM}(t)=S_{DM}\left(t_0\right)\exp\left({\int_{t_0}^{t}{\Gamma _2\frac{\rho _{\phi }}{n_{DM}}dt}}\right).
\ee

Irreversible matter particle creation is an adiabatic process, the produced entropy being entirely due to the increase in the number of fluid particles, there being no increase in the entropy per particle due to dissipative processes.

\section{Irreversible dark energy-dark matter interaction models}\label{sect4}

In the present Section we consider, within the framework of irreversible thermodynamics with matter creation/annihilation, a number of specific cosmological models with dark energy-dark matter interaction. As a first case we consider the situation in which the density of the dark matter is much smaller than the energy density of the scalar field. This case corresponds to an Universe dominated by the dark energy component, assumed to be represented by a coherent wave of $\phi$-particles. In this case the kinetic term dominates in the total energy of the scalar field. As a second case we consider a potential energy dominated scalar field. The general dynamics of the cosmological model with interacting dark energy and dark matter is also considered, and the cosmological evolution equations are studied numerically.

\subsection{Coherent Scalar Waves-Dark matter interaction}

We shall consider in the following that the energy density and particle number of the newly created dark  matter is much smaller than the energy density
and particle number of the corresponding scalar field fluid component, that is the relations $\rho _{DM} \ll \rho _{\phi }$, $n_{DM} \ll n_{\phi }$, and $p_{DM}  \ll p_{\phi }$ hold. In this case the Universe is dominated by the scalar field energy density, and its evolution is not influenced
by the matter content. We shall work throughout with finite values of the fluid quantities at $t = t_0$. The coupling between scalar field and dark matter is realised only by means of the balance equation of ordinary matter via the scalar field energy density, and the basic equations describing the dynamics of a flat FRW scalar field filled space-time interacting with a dark matter component are given by
\bea
3H^2&=&\rho _{\phi }
   \\
%\ee
%\be
2\dot{H}+3H^2&=&-p_{\phi }-\frac{\Gamma _1\left(\rho _{\phi }+p_{\phi }\right)\rho _{\phi }}{3Hn_{\phi}},
\eea
and
\be \label{45}
\dot{\rho} _{\phi}=\left(\rho _{\phi }+p_{\phi }\right)\frac{\dot{n}_{\phi }}{n_{\phi }},
\ee
\bea
\dot{n}_{\phi }+3Hn_{\phi }&=&-\Gamma _1\rho _{\phi },
%\ee
%\be
   \\
\dot{n}_{DM }+3Hn_{DM }&=&\Gamma _2\rho _{\phi },
\eea
respectively.

A homogeneous scalar field oscillating with frequency $m_{\phi }$  can be considered as a coherent wave of ``particles'' with zero momenta, and with a particle number density given by \cite{Linde}
\be\label{bar1}
n_{\phi }=\frac{\rho _{\phi}}{m_{\phi }},\qquad m_{\phi }={\rm constant}.
\ee
In other words, $n_{\phi}$ oscillators of the same frequency $m_{\phi }$ oscillating coherently with the same phase can be described as a single homogeneous wave $\phi (t)$. Insertion of the energy density of the scalar field given by Eq.~(\ref{bar1}) in
Eq.~(\ref{45}) leads to the condition
\be
p_{\phi }=0,
\ee
or, equivalently,
\be
U\left(\dot{\phi }\right)=\frac{\dot{\phi }^2}{2}.
\ee
Therefore, a homogeneous oscillating scalar field is described in the present model by a Barrow-Saich
type potential, with the potential energy of the scalar field proportional
to the kinetic one \cite{Bar}. The energy density of the scalar field
becomes $\rho _{\phi } = \dot{\phi }^2$,  and this relation, obtained naturally in the framework of the present
formalism is very similar to the equation $\rho _{\phi }=\left<\dot{\phi }^2\right> $
obtained by replacing $\dot{\phi }^2$ by its average value per cycle \cite{Kolb}.

In this case the equations describing the dynamics of the FRW type spacetime
filled by the decaying oscillating homogeneous scalar field in the presence of
dark matter creation become
\be
3H^2 =\dot{\phi }^2,
\ee
\be\label{52}
2\dot{H}+3H^2=-\Gamma _1m_{\phi} H,
\ee
\be\label{53}
\dot{n}_{DM}+3Hn_{DM}=3\Gamma _2H^2,
\ee
respectively.
In the following we will assume, for simplicity, that $\Gamma _1$ and $\Gamma _2$ are constants.  By introducing a set of dimensionless variables $\tau $, $h$ and $\theta _{DM}$ by means of the
transformations
\be
t=\frac{2 }{\Gamma _1m_{\phi }}\tau,\quad H=\frac{\Gamma _1m_{\phi }}{3}h, \quad n_{DM}=\frac{2\Gamma _1\Gamma _2m_{\phi }}{3}\theta _{DM},
\ee
Equations (\ref{52}) and (\ref{53}) take the form
\be
\frac{dh}{d\tau }=-h(h+1),
\ee
\be
\frac{d\theta _{DM}}{d\tau }+2h\theta _{DM}=h^2,
\ee
and yield the following general solutions
\be
h(\tau)=\frac{1}{e^{\tau }-1},
\ee
and
\be
\theta _{DM}=\frac{\left[\left(e^{\tau _0}-1\right)^2\theta _{DM0}+1/2\right]e^{2\left(\tau -\tau _0\right)}-1/2}{\left(e^{\tau }-1\right)^2},
\ee
respectively, where we have denoted $\theta _{DM0}= \theta _{DM}\left(\tau _0\right)$.

The evolution of the scale factor is given by
\be
a(\tau )=a_0\left(\frac{e^{\tau }-1}{e^{\tau }}\right)^{2/3}.
\ee
The deceleration parameter $q=d(1/H)/dt-1$ is given by
\be
q=\frac{3}{2}e^{\tau }-1.
\ee
For $\left(t- t_0\right) \ll \Gamma _1^{-1}$, i.e., at the start of the oscillatory period corresponding to
the dark matter production, the approximate solution of the field equations is
given by
\be
h\approx \frac{1}{\tau }, \qquad  a\approx a_0\tau ^{2/3},
\ee
\be
\rho _{\phi }\approx \frac{1}{\tau ^2}, \qquad   \theta _{DM}\approx \theta _{DM0}\frac{\tau _0^2}{\tau ^2}\left(1+\frac{\tau -\tau _0}{\theta _{DM0}\tau _0^2}\right).
\ee
This phase corresponds to an Einstein-de Sitter expansion, with decaying dark energy, and dark matter creation.

During the initial oscillating period of the scalar field dominated FRW Universe there is a rapid increase of its dark matter content. The particle number density increases during a time interval
\be
\Delta t = t_{max} -t_0 = \left(1-\frac{3\theta _{DM0}t_0}{2\Gamma _2}\right)t_0,
\ee
and reaches a maximum value given by
\be
n_{DM}^{(max)}=\frac{\Gamma _2}{3\Delta t}\frac{1-3\theta _{DM0}t_0/2\Gamma _2}{t_0\left(1-3\theta _{DM0}t_0/4\Gamma _2\right)}.
\ee

Assuming that the  scalar field oscillations decay into relativistic dark matter, the energy density and temperature of the dark matter component of
the cosmological fluid is given by $\rho _{DM}\sim n^{\gamma }$, $T_{DM}\sim n^{\gamma -1}$, where $\gamma = 2$ corresponds
to a stiff (Zeldovich) fluid obeying an equation of state of the form $\rho _{DM} = p_{DM}$, and $\gamma = 4/3$
corresponds to a relativistic, radiation like  fluid.

The entropy produced during dark matter particle creation can be easily obtained from Eq.~(\ref{42}) and is given, in first approximation, by
\be
\frac{S_{DM}(t)}{S_{DM0}}=\frac{\tau-\tau _0}{\theta _{DM0}\tau _0^2}+1.
\ee
Therefore, for small times, there is a linear increase of the entropy of the dark energy dominated flat spacetime.

\subsection{Constant scalar field potential dark energy and dark matter  interaction}

As a second example in the study of the irreversible interaction between dark energy (a scalar field) and dark matter we consider the case in which the scalar field potential may be approximated, at least for a certain time interval, as a positive constant, $U(\phi )=\Lambda ={\rm constant}>0$. Therefore, the energy density and the pressure of the scalar field can be written as
\be
\rho _{\phi }=\frac{1}{2}\dot{\phi }^2+\Lambda, \qquad \rho _{\phi }=\frac{1}{2}\dot{\phi }^2-\Lambda .
\ee

Moreover, we assume that the change in the scalar field energy density due to the cosmological expansion can be neglected in the scalar field evolution equations, that is, in the energy density and particle number equations of the scalar field, given by Eqs.~(\ref{scalf}) and (\ref{dm}), the term containing $3H$  can be neglected. In this approximation the main contribution to the temporal dynamics of the scalar field is due to its decay into dark matter particles, and not to the cosmological expansion. On the other hand, the feedback of the newly created dark matter particles on the cosmological dynamics cannot be neglected.
Within this approximation, from Eq.~(\ref{rate1}), describing the scalar field particles decay,  we obtain first
\be
\rho _{\phi}=-\frac{1}{\Gamma _1}\dot{n}_{\phi},
\ee
which gives the scalar field particle number density as a function of the energy of the scalar field. By substituting this expression of $\rho _{\phi }$ into Eq.~(\ref{scalf}) gives
\be
\ddot{n}_{\phi }+\Gamma _1\frac{\dot{\phi}^2}{n_{\phi }}\dot{n}_{\phi }=0.
\ee
Taking into account that
\be
\dot{\phi }^2=2\left(\rho _{\phi }-\Lambda\right)=2\left(-
\frac{\dot{n}_{\phi }}{\Gamma_1}-\Lambda \right) \,,
\ee
we obtain the equation describing the dynamics of the scalar field particles as
\be\label{eqn}
n_{\phi}\ddot{n}_{\phi}-2n_{\phi }^2-2\Gamma _1\Lambda \dot {n}_{\phi}=0.
\ee
By denoting $\dot{n}_{\phi }=u$, $\ddot{n}_{\phi }=udu/dn_{\phi }$, Eq.~(\ref{eqn}) becomes
\be
n_{\phi}\frac{du}{dn_{\phi }}=2\left(u+\Gamma _1\Lambda \right),
\ee
providing
\be
\dot{n}_{\phi}=N_1n_{\phi }^2-\Gamma _1\Lambda,
\ee
where $N_1$ is an arbitrary integration constant, which can be determined from the initial condition $\dot{n}_{\phi }\left(t_0\right)=-\Gamma _1\rho _{\phi }\left(t_0\right)=-\Gamma _1\rho _{\phi 0}$ and $n_{\phi}\left(t_0\right)=n_{\phi 0}$ as
\be
N_1=\frac{\Gamma _1\left(\Lambda -\rho _{\phi 0}\right)}{n_{\phi 0}^2}.
\ee
Therefore the general solution of Eq.~(\ref{eqn}) is given by
\be
n_{\phi }(t)=\sqrt{\frac{\Gamma _1\Lambda }{N_1}}\frac{1+N_2e^{2\alpha t}}{1-N_2e^{2\alpha t}},
\ee
where $\alpha =\sqrt{\Gamma _1\Lambda N_1}$, and
\be
N_2=\frac{n_{\phi 0}-\sqrt{\Gamma _1\Lambda /N_1}}{n_{\phi 0}+\sqrt{\Gamma _1\Lambda /N_1}}e^{-2\alpha t_0}.
\ee

By assuming that the newly created dark matter particles are pressureless, $p_{DM }=0$, and the energy density of the dark matter component is $\rho _{DM}=m_{DM}n_{DM}$, where $m_{DM}$ is the mass of the dark matter particle, then the evolution equation for the dark matter density can be written as
\bea
\dot{\rho }_{DM}+3H\rho _{DM}&=&-\frac{\Gamma _2}{\Gamma _1}\dot{n}_{\phi }
   \nonumber  \\
&=&\Gamma _2\Lambda -\frac{\Gamma _2\left(\Lambda -\rho _{\phi 0}\right)}{n_{\phi 0}^2}n_{\phi }^2,
\eea
By neglecting the effects of the expansion of the Universe the dark matter energy increases as
\bea
\rho _{DM}(t)&\approx &\Gamma _2\left(\Lambda -\sqrt{\frac{\Gamma _1\lambda }{N_1}}\frac{\Lambda -\rho _{\phi 0}}{n_{\phi 0}^2}\right)t \nonumber\\
&&- \frac{\Gamma _2\left(\Lambda -\rho _{\phi 0}\right)}{\alpha n_{\phi 0}^2}\sqrt{\frac{\Gamma _1\lambda }{N_1}}\frac{1}{N_2e^{2\alpha t}-1}.
\eea

\subsection{Effects of the cosmological expansion on  the dark energy-dark matter interaction}

In the general case, in which the expansion of the Universe is also taken into account, the system of equations describing the irreversible dark energy-dark matter interaction are given by
\be\label{f1}
3H^2=\rho _{\phi }+\rho _{DM},
\ee
\be
\dot{n}_{\phi }+3Hn_{\phi }=-\Gamma _1\rho _{\phi },
\ee
\be
\dot{\rho} _{\phi }+6H\left(\rho _{\phi }-\Lambda \right)+\frac{2\Gamma _1\left(\rho _{\phi }-\Lambda \right)\rho _{\phi }}{n_{\phi }}=0,
\ee
\be\label{f4}
\dot{\rho }_{DM}+   3H\rho _{DM}=\Gamma _1m_{DM}\rho _{\phi },
\ee
where, for simplicity, we have assumed $\Gamma _1=\Gamma _2={\rm constant}$. By introducing a set of dimensionless variables $\left(r_{\phi}, N_{\phi },r_{DM}, \tau\right)$, defined as
\bea
\rho _{\phi}=\Lambda r_{\phi },\quad n_{\phi}=\Lambda N_{\phi },
     \\
\rho _{DM}=\Lambda m_{DM}r_{DM}, \quad t=\tau/\Gamma _1,
\eea
the field equations Eqs.~(\ref{f1})-(\ref{f4}) can be written in a dimensionless form as
\be
\frac{1}{a}\frac{da}{d\tau}=\lambda \sqrt{r_{\phi }+m_{DM}r_{DM}},
\ee
\be
\frac{dN_{\phi }}{d\tau }+3\lambda \sqrt{r_{\phi }+m_{DM}r_{DM}}N_{\phi }=-r_{\phi },
\ee
\be
\frac{dr_{\phi}}{d\tau }+6\lambda \sqrt{r_{\phi }+m_{DM}r_{DM}}\left(r_{\phi }-1\right)+\frac{2r_{\phi }\left(r_{\phi }-1\right)}{N_{\phi }}=0,
\ee
\be
\frac{dr_{DM}}{d\tau}+3\lambda \sqrt{r_{\phi }+m_{DM}r_{DM}}r_{DM}=r_{\phi },
\ee
respectively, where
\be
\lambda =\frac{\sqrt{\Lambda /3}}{\Gamma _1}.
\ee

The deceleration parameter $q$ in the irreversible interacting dark energy-dark matter model is given by
\begin{widetext}
\be
q=-\frac{1}{2}\frac{r_{\phi}\left[m_{DM}-2\left(r_{\phi}-1\right)/N_{\phi }\right]-3\lambda \sqrt{r_{\phi }+m_{DM}r_{DM}}\left[2\left(r_{\phi }-1\right)+m_{DM}r_{DM}\right]}{\lambda \left(r_{\phi }+m_{DM}r_{DM}\right)^{3/2}}-1.
\ee
\end{widetext}

The density parameters $\Omega _{\phi}$ and $\Omega _{DM}$ of the dark energy and of the dark matter are given by
\be
\Omega _{\phi }=\frac{r_{\phi }}{r_{\phi}+m_{DM}r_{DM}},
\ee
and
\be
\Omega _{DM}=\frac{m_{DM}r_{DM}}{r_{\phi }+m_{DM}r_{DM}},
\ee
respectively, and they satisfy the relation $\Omega _{\phi }+ \Omega _{DM}=1$.

The dynamics of the interacting dark energy-dark matter system is determined by two control parameters, $\lambda $, and the mass $m_{DM}$ of the dark matter particle. The mass of the dark matter particle, as well as the decoupling temperature, was determined recently in \cite{Sal1}.  By evaluating
analytically the dark matter galaxy properties, as the halo density profile, the halo radius and the surface density, and by matching them to their observed
values one can obtain the decreasing of the phase space density since equilibration till today, the mass of the dark matter particle and the decoupling temperature $T_d$, and the kind of the halo density profile (core or cusp), respectively. The dark matter particle mass turns to be between 1 and 2 keV and the decoupling temperature $T_d$ turns to be above 100 GeV. Dark matter particles with masses of the order of keV  necessarily produce cored density profiles, while WIMPs ($m\sim  100$ GeV, $T_d\sim 5$ GeV) inevitably produce cusped profiles at scales about 0.003 pc. Therefore, based on this analysis in the following we adopt a value of 1 keV for the mass of the dark matter particle.  Some possible dark matter candidates with masses in this range would be the sterile neutrino, the gravitino, the light neutralino, the majoron etc. \cite{Sal1}.

By normalizing the value of the scale factor so that at the present time $t_{pres}$  the scale factor is $a\left(t_{pres}\right)=1$, the redshift $z$ is given by $z=(1-a)/a$. We consider the effects of the dark energy-dark matter interaction in the recent Universe, starting from $z=2$, corresponding to $t=0$, which fixes the initial value of the scale factor as $a(0)=0.33$. Moreover, we consider that at $z=2$ the universe was composed of an equal amount of dark energy and dark matter so that $r_{\phi}(0)=r_{DM}(0)$, and for the numerical calculations we choose $r_{\phi}(0)=r_{DM}(0)=1.5$.

\begin{figure}
 \centering
 \includegraphics[scale=0.70]{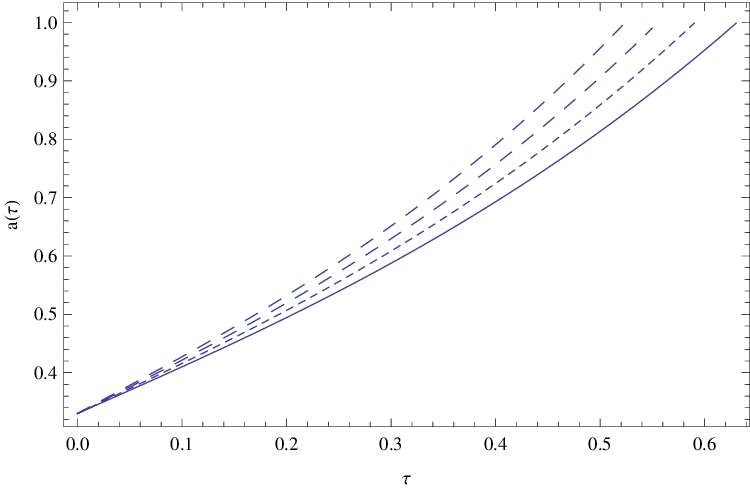}
 \caption{Time variation of the scale factor of the interacting dark
 energy-dark matter filled Universe, for different values of the parameter
 $\lambda $: $\lambda =1.4$ (solid curve), $\lambda =1.5$ (dotted curve),
 $\lambda =1.6$ (dashed curve), and $\lambda =1.7$ (long dashed curve).
 The initial conditions used for the numerical integration of the cosmological
 evolution equations are $a(0)=0.33$, $N_{\phi }(0)=10$, $r_{\phi }(0)=1.5$,
 and $r_{DM}(0)=1.5$, respectively. For the mass $m_{DM}$ of the dark matter
 particle we have assumed the value $m_{DM}=1$ keV.}
 \label{a}
\end{figure}

\begin{figure}
 \centering
 \includegraphics[scale=0.70]{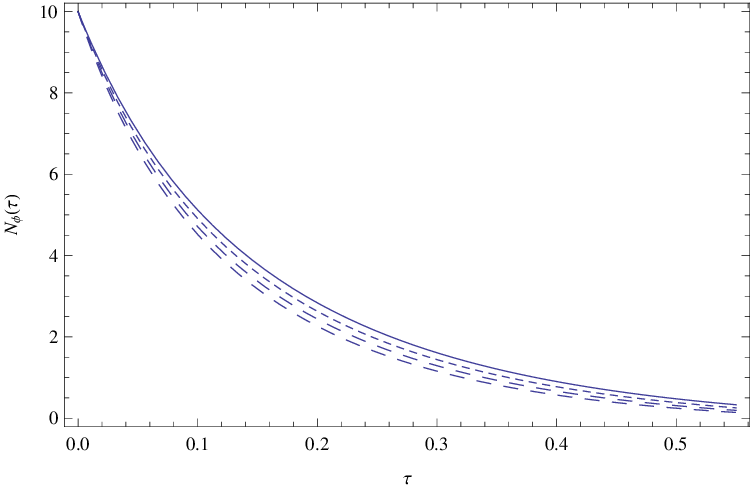}
 \caption{Time variation of the scalar field particle number $N_{\phi}$ in the
 interacting dark energy-dark matter filled Universe, for different values of
 the parameter $\lambda $: $\lambda =1.4$ (solid curve), $\lambda =1.5$ (dotted curve),
 $\lambda =1.6$ (dashed curve), and $\lambda =1.7$ (long dashed curve). The initial conditions used for the numerical integration of
 the cosmological evolution equations are $a(0)=0.33$, $N_{\phi }(0)=10$,
 $r_{\phi }(0)=1.5$, and $r_{DM}(0)=1.5$, respectively. For the mass $m_{DM}$
 of the dark matter particle we have assumed the value $m_{DM}=1$ keV.}
 \label{N}
\end{figure}

\begin{figure}
 \centering
 \includegraphics[scale=0.70]{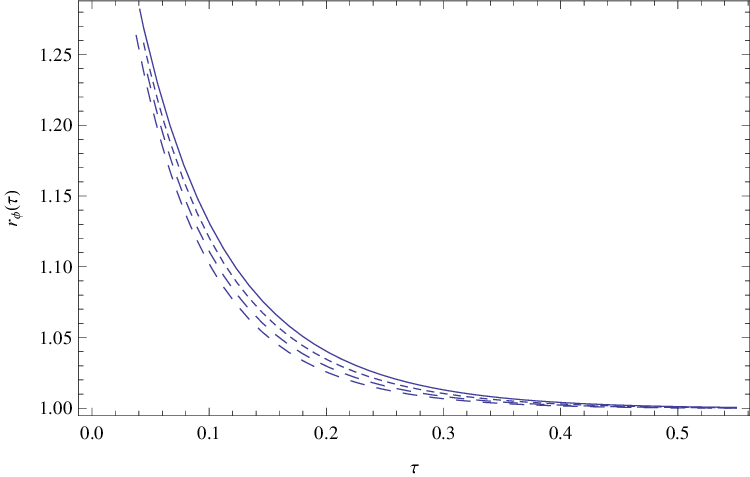}
 \caption{Time variation of the dimensionless scalar field energy $r_{\phi}$
 in the interacting dark energy-dark matter filled Universe, for different
 values of the parameter $\lambda $: $\lambda =1.4$ (solid curve), $\lambda =1.5$ (dotted curve),
 $\lambda =1.6$ (dashed curve), and $\lambda =1.7$ (long dashed curve). The initial conditions used for the numerical
 integration of the cosmological evolution equations are $a(0)=0.33$, $N_{\phi
 }(0)=10$, $r_{\phi }(0)=1.5$,  and $r_{DM}(0)=1.5$, respectively. For the
 mass $m_{DM}$ of the dark matter particle we have assumed the value $m_{DM}=1$
 keV.}
 \label{rf}
\end{figure}

\begin{figure}
 \centering
 \includegraphics[scale=0.70]{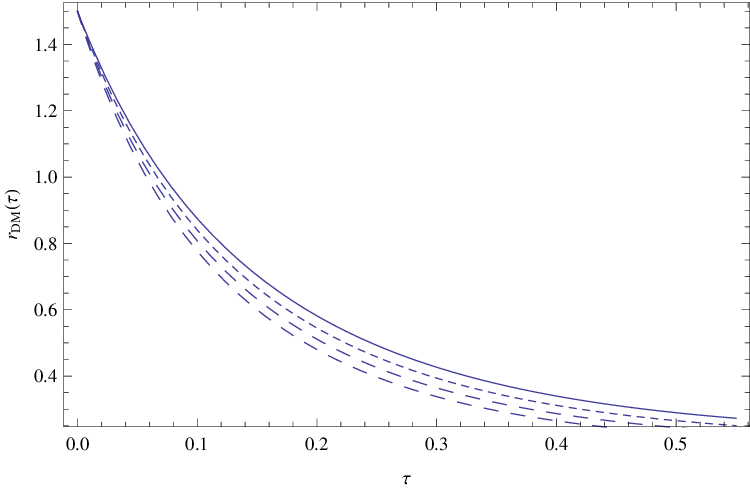}
 \caption{Time variation of the dimensionless dark matter  energy $r_{DM}$ in
 the interacting dark energy-dark matter filled Universe, for different values
 of the parameter $\lambda $:$\lambda =1.4$ (solid curve), $\lambda =1.5$ (dotted curve),
 $\lambda =1.6$ (dashed curve), and $\lambda =1.7$ (long dashed curve). The initial conditions used for the numerical integration of
 the cosmological evolution equations are $a(0)=0.33$, $N_{\phi }(0)=10$,
 $r_{\phi }(0)=1.5$,  and $r_{DM}(0)=1.5$, respectively. For the mass $m_{DM}$
 of the dark matter particle we have assumed the value $m_{DM}=1$ keV.}\label{fig4}
\end{figure}

\begin{figure}
 \centering
 \includegraphics[scale=0.70]{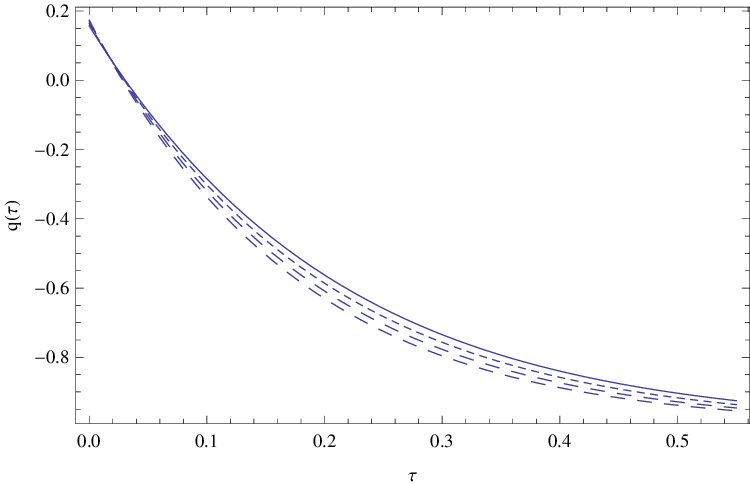}
 \caption{Time variation of the deceleration parameter $q$ in
 the interacting dark energy-dark matter filled Universe, for different values
 of the parameter $\lambda $: $\lambda =1.4$ (solid curve), $\lambda =1.5$ (dotted curve),
 $\lambda =1.6$ (dashed curve), and $\lambda =1.7$ (long dashed curve). The initial conditions used for the numerical integration of
 the cosmological evolution equations are $a(0)=0.33$, $N_{\phi }(0)=10$,
 $r_{\phi }(0)=1.5$, and $r_{DM}(0)=1.5$, respectively. For the mass $m_{DM}$
 of the dark matter particle we have assumed the value $m_{DM}=1$ keV.}\label{fig5}
\end{figure}

\begin{figure}
 \centering
 \includegraphics[scale=0.70]{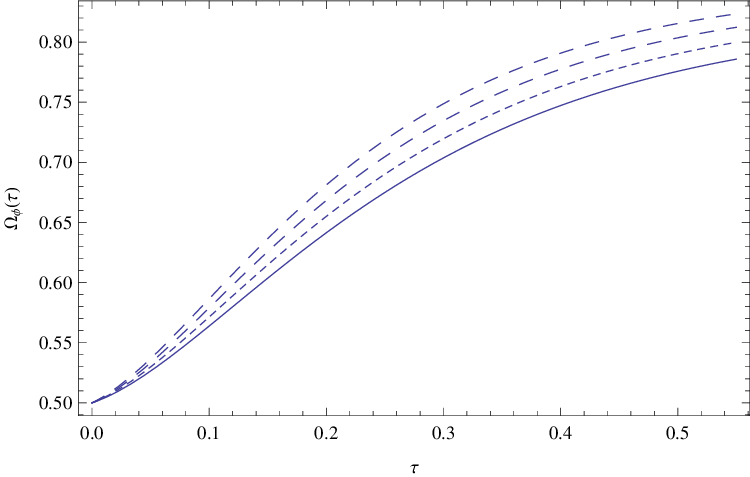}
 \caption{Time variation of the density parameter $\Omega _{\phi}$ of the dark energy   in
 the interacting dark energy-dark matter filled Universe, for different values
 of the parameter $\lambda $: $\lambda =1.4$ (solid curve), $\lambda =1.5$ (dotted curve),
 $\lambda =1.6$ (dashed curve), and $\lambda =1.7$ (long dashed curve). The initial conditions used for the numerical integration of
 the cosmological evolution equations are $a(0)=0.33$, $N_{\phi }(0)=10$,
 $r_{\phi }(0)=1.5$,  and $r_{DM}(0)=1.5$, respectively. For the mass $m_{DM}$
 of the dark matter particle we have assumed the value $m_{DM}=1$ keV.}\label{fig6}
\end{figure}

\begin{figure}
 \centering
 \includegraphics[scale=0.70]{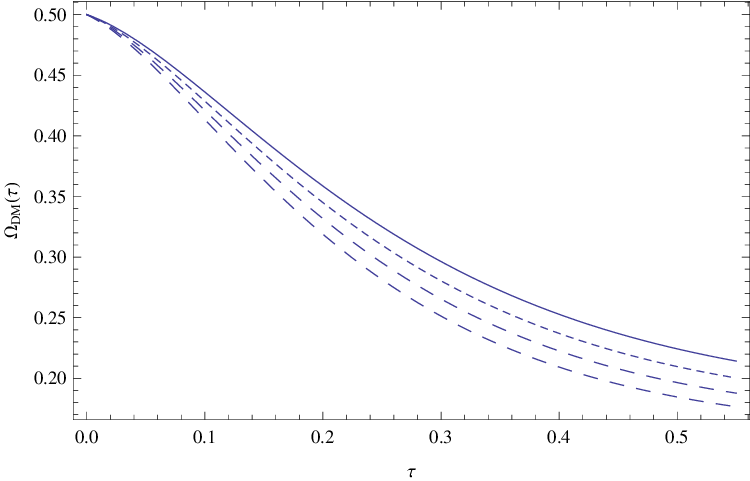}
 \caption{Time variation of the density parameter $\Omega _{DM}$ of the dark matter    in
 the interacting dark energy-dark matter filled Universe, for different values
 of the parameter $\lambda $: $\lambda =1.4$ (solid curve), $\lambda =1.5$ (dotted curve),
 $\lambda =1.6$ (dashed curve), and $\lambda =1.7$ (long dashed curve). The initial conditions used for the numerical integration of
 the cosmological evolution equations are $a(0)=0.33$, $N_{\phi }(0)=10$,
 $r_{\phi }(0)=1.5$,  and $r_{DM}(0)=1.5$, respectively. For the mass $m_{DM}$
 of the dark matter particle we have assumed the value $m_{DM}=1$ keV.}\label{fig7}
\end{figure}

The time variations of the scale factor $a$, of the scalar field particle number $N_{\phi }$, of the scalar field energy $r_{\phi }$, of the dark matter energy density $r_{DM}$, of the deceleration parameter $q$, of the density parameter $\Omega _{\phi}$, and of the density parameter $\Omega _{DM}$ of the dark matter  are represented, for different values of the parameter $\lambda $, and for $m_{DM}=1$ keV,  in Figs.~\ref{a}-\ref{fig7}.

As one can see from Fig.~\ref{a}, the interacting dark energy-dark matter filled Universe is an expansionary state, with the rate of the expansion, and the scale factor evolution, strongly dependent on the dimensionless parameter $\lambda $, which depends on the ratio of the (constant) scalar field potential $\Lambda $, and of the scalar field decay rate $\Gamma _1$. Accelerated expansion can also be obtained in the framework of the present model, the creation pressure, corresponding to the irreversible decay of the scalar field, and the matter creation, can drive the Universe into a de Sitter type phase. The scalar field particle number decays during the cosmological evolution, as shown in Fig.~\ref{N}. The decay rate strongly depends on the numerical value of $\lambda $. The dimensionless energy of the scalar field $r_{\phi }$, shown in Fig.~\ref{rf}, tends in the large time limit to the value 1, corresponding to $\rho _{\phi }=\Lambda$, and to a de Sitter type expansion. This shows that the decay of the scalar field is determined and controlled by the kinetic energy term of the field $\dot{\phi }^2/2$, which is the source of the dark matter particles creation. When the energy and the pressure of the scalar field are dominated by the scalar field potential $\Lambda $, $\rho _{\phi}=-p_{\phi }=\Lambda $, then $\rho _{\phi }+p_{\phi }=0$, and from Eq.~(\ref{scalf}) it follows that $\rho _{\phi}={\rm constant}$, and the scalar field energy cannot be converted any more into other type of particles. The energy density of the dark matter particles, presented in Fig.~\ref{fig4}, increases in time due to the decay of the $\phi $-particles.

The time variation of the deceleration parameter $q$, presented in Fig.~\ref{fig5}, shows that the Universe with irreversibly interacting dark energy-dark matter starts its evolution at $z=2$ from a decelerating state, with $q>0$, and with an initial value of the decelerating parameter of around $q(0)\approx 0.2$. For the present choice of the parameters the Universe enters into an accelerating phase, with $q<0$, at a redshift of around $z\approx 1.6$, and reaches  a de Sitter type expansionary phase at the present time, corresponding to $a=1$, and $z=0$, respectively.

The time variations of the density parameters of the dark energy and of the dark matter are represented, for different values of $\lambda $, in Figs.~\ref{fig6} and \ref{fig7}, respectively. At $t=0$ ($z=2$) dark energy and dark matter have the same values of the density parameters, $\Omega _{\phi }(0)=\Omega _{DM}(0)=1/2$. During the cosmological expansion in the redshift range $2\leq z\leq 0$, the density parameter of the dark energy increases to a value of $\Omega _{\phi }\left(t_{pres}\right)\approx 0.8$, while at the same time the density parameter of the dark matter decreases to $\Omega _{\phi }\left(t_{pres}\right)\approx 0.2$. For $\lambda =1.4$, $\Omega _{\phi}\approx 0.77$, and $\Omega _{DM}\approx 0.23$. These results are consistent with the latest observational determinations of the composition of the Universe, which give $\Omega _{\phi }\left(t_{pres}\right)\approx 0.73$, and $\Omega _{DM }\left(t_{pres}\right)\approx 0.228$ \cite{Hin}.

\section{Discussions and final remarks}\label{sect5}

In the present paper we have shown that the thermodynamics of the open systems is a valuable tool for describing the interacting dark energy - dark matter  phases of the general relativistic cosmological models. As applied to a two component (scalar field and dark matter) cosmological model, the thermodynamics of irreversible processes provides a generalization of the elementary theory of dark energy-dark matter interaction, which envisages that during the dark energy dominated phase of the expansion of the Universe, when the expansion slows down, the energy stored in the zero mode oscillations  of the scalar field transforms into particles via single particle decay. Thus the model presented in this paper gives a rigorous thermodynamical foundation, and a natural generalisation, to the theory of dark energy-dark matter interaction. Particle decay (creation) gives rise to a supplementary decay (creation) pressure which has to be included as a distinct part in the stress-energy tensor of the cosmological mixture. We have considered only particle creation in a scalar field (dark energy) dominated Universes, leading to a model in which particle production rate is extremely high at the beginning of the oscillatory period, and afterwards it tends to zero, when the scalar field energy and pressure become dominated by the scalar field potential, assumed to be a constant, and when the kinetic energy of the scalar field becomes negligibly small. Such a dark matter particle production has very important implications on the dynamics and evolution of  the Universe. The details of the dark energy-dark matter interaction  mechanism depend on the parameters of the particle physics models involved to describe the newly created particles. One of the parameters is the dark matter particle mass, which is a key parameter in the description of the scalar field-dark matter process. Unfortunately presently there is no definite answer giving the value of the mass of the dark matter particle.

Particle creation can be related to what is called the arrow of time: something that provides a direction to time, and distinguishes the future from the past. There are two different arrows of time: the thermodynamical arrow of time, the direction in which entropy increases, and the cosmological arrow of time, the direction in which the Universe is expanding. Particle creation introduces asymmetry in the evolution of the Universe, and enables us to assign a thermodynamical arrow of time, which agrees, in our model, with the cosmological one. This coincidence is determined in a natural way by the decay of the scalar field, due to the presence of a friction force between field and matter.

In our present approach we have neglected the back-reaction of the newly created particles on the dynamics of the Universe, and the effect of the form of the scalar field potential $U(\phi )$ have not been completely envisaged. These aspects of the thermodynamic theory of the dark energy - dark matter interaction will be the subject of a future work.

\section*{Acknowledgments}
 FSNL acknowledges financial support of the Funda\c{c}\~{a}o para a Ci\^{e}ncia e Tecnologia through the grants CERN/FP/123615/2011 and CERN/FP/123618/2011.

\end{document}